\documentstyle[12pt,epsf]{article}
\newcommand{\beq}{\begin{equation}}
\newcommand{\eeq}{\end{equation}}
\newcommand{\bea}{\begin{eqnarray}}
\newcommand{\eea}{\end{eqnarray}}
\newcommand{\eps}{\varepsilon}
\newcommand{\Veff}%
{{\cal V}^{\mbox{\scriptsize p}}_{\mbox{\scriptsize eff}}}

\begin{document}

\centerline{\large\bf A simple model for the microscopic effective }
\centerline{\large\bf pairing interaction}

\vskip 1 cm

\centerline{M.$\,$Baldo$^{1}$, U.$\,$Lombardo$^{2,3}$,
E.$\,$Saperstein$^{4}$ and M.$\,$Zverev$^{4}$}

\vskip 1 cm

\centerline{$^1$INFN, Sezione di Catania, 57 Corso Italia,
I-95129 Catania, Italy}
\centerline{$^2$INFN-LNS, 44 Via S.-Sofia, I-95123 Catania, Italy}
\centerline{$^3$ Dipartimento di Fisica, 57 Corso Italia,
I-95129 Catania, Italy}

\centerline{$^4$ Kurchatov Institute, 123182, Moscow, Russia }

\vskip 3 cm

\centerline
{\bf Abstract}
\vskip .3 cm

The microscopic effective pairing interaction in the $^1S_0$-channel
is investigated for two different
values of the chemical potential $\mu$ starting from the separable
form of the Paris NN-potential.
It is shown that, within a high accuracy, this effective interaction
can be approximated by the off-shell free $T$-matrix taken 
at the negative energy $E=2\mu$.   

\newpage

Recently  \cite{BLSZ1} the microscopic effective pairing interaction 
$\Veff$ in the $^1S_0$ -channel  was found by solving
the Bethe-Goldstone equation for semi-infinite nuclear matter
without any form of local approximation.              
The separable representation  \cite{Par1,Par2} of the 
Paris potential \cite{Paris} was used which simplifies the
problem significantly and makes it possible such direct
solution for non-uniform systems with the use of the
mixed coordinate-momentum representation.
However, the procedure turned out to be very cumbersome
and the effective interaction obtained has rather complicated
form not convenient for applications.
In this paper we show that $\Veff$, with a good accuracy,
can be approximated by the free $T$-matrix taken at the
energy $E=2\mu$. This quantity is much simpler than    
$\Veff$ and can be easily found for a fixed value of $\mu$.

The many-body theory form of the gap equation \cite{AB,Schuck} 
\begin{equation}
\Delta = {\cal V} G G^s \Delta
\end{equation}
is used in \cite{BLSZ1}
which explicitly takes into consideration the particle-particle propagator
$ A^s = G G^s $ in the superfluid system. Within the Bethe-Brueckner
approach, the irreducible particle-particle interaction
block ${\cal V}$ is approximated by the free $NN$-potential.
The effective interaction is associated with the splitting of the complete 
Hilbert space $S$ into two
domains, the model subspace $S_0$ and the complementary one $S'$. 
As a result, the two-particle propagator
is represented as a sum $A^s = A^s_0 + A'$. 
It is supposed that the superfluid effects can be neglected in the 
$S'$-subspace and therefore the upperscript ``s'' is omitted
in the second term.

The gap equation (1) can be rewritten in the model subspace:
\begin{equation}
\Delta = {\Veff} A^s_0 \Delta,
\end{equation}
where $\Veff$ obeys the following equation:
\begin{equation}
{\Veff} = {\cal V} + {\cal V} A^{\prime}{\Veff}.
\end{equation}

In \cite{BLSZ1} the model space $S_0$ is defined in such a way that
it involves all the two-particle states $(\lambda ,\lambda')$
with the single-particle energies $\eps_{\lambda}, \eps_{\lambda'} $ both 
negative. 
In this case, the complementary subspace $S'$ involves all the 
two-particle states
 with  positive energies $  \eps_{\lambda},  \eps_{\lambda^{\prime}} $
and two-particle states with one energy positive and the second one 
negative (but greater than $\mu$).

The separable form \cite{Par1,Par2} of the
Paris potential is as follows:
\begin{equation}
{\cal V}({ k},{ k}^{\prime}) =
\sum_{ij} \lambda_{ij} g_i(k^2) g_j(k^{\prime 2}),
\end{equation}
where form factors $g_i(k^2),\; (i=1,2,3)$ are rational functions. 
This choice was adopted earlier 
in the Brueckner-type calculations for infinite nuclear matter \cite{BL1,BL2}.
It should be
noted that the original normalization  \cite{Par1,Par2} of the
expansion (4) was changed in \cite{BLSZ1} in such a way that
the identity  $g_i(0)=1$ holds true. Then the absolute values
of the $\lambda_{ij}$-coefficients give direct information
on the strength of the corresponding terms of the force.
Their values (in MeV$\cdot$fm$^3$) are as follows:
$\lambda_{11}{=} -3659,\; \lambda_{12} {=}  2169,\; \lambda_{22} {=} -1485$
 and $\lambda_{13}{=} -23.6,\;  \lambda_{23}{=} 57.6,\; \lambda_{33}{=} 17.2$. 
As it is seen, the strengths of all the components
containing only the indices $i=1,2$ are
much stronger than those with the index $i=3$. Therefore the
latter will not be considered for a qualitative analysis. Of course,
in the calculations all the terms $\lambda_{ik}$
are considered. 

The separable representation of the $NN$-potential leads to a similar form
of $\Veff$ which, in the notation of \cite{BLSZ1} is as follows:
\begin{equation}
{\Veff}(k^2_{\perp},k^{\prime 2}_{\perp};x_1,x_2,x_3,x_4;E) =
\sum\limits_{ij}\Lambda_{ij}(X_{12},X_{34};E)
g_i(k^2_{\perp},x_{12}) g_j(k^{\prime 2}_{\perp},x_{34}).
\end{equation}
Here the center-of-mass and relative coordinates
in the $x$-direction are introduced
($X_{12}=(x_1{+}x_2)/2$, $x_{12}=x_1{-}x_2$, etc.), and $g_i(k^2_{\perp},x)$ 
stands for the inverse Fourier transform of
the form factors $g_i(k^2_{\perp}+k^2_x)$ in the $x$-direction.

The coefficients $\Lambda_{ij}$ obey the set of
integral equations:
\begin{eqnarray}
\lefteqn{
\Lambda_{ij} (X_{12},X_{34};E) =\lambda_{ij} \delta(X_{12}{-}X_{34}) +
}
\nonumber \\
& &{} + \sum_{lm}\lambda_{il}\int dX_{56}\,B_{lm}(X_{12},X_{56};E)
\,\Lambda_{mj}(X_{56},X_{34};E),
\end{eqnarray}
where $B_{lm}$ are given by integrals of the propagator $A'$ 
with two form factors. Their explicit form is as follows: 
\beq
B_{lm}(X_{12},X_{34};E) = { \sum_{nn'}}^{\prime}
           \int{ d^2{\bf k}_{\perp} \over {(2\pi)^2} }
\frac   
{G_{nn^{\prime}}^l(k^2_{\perp},X_{12})\,
G_{n^{\prime}n}^{m*}(k^2_{\perp},X_{34})}
{E -\eps_{\lambda} -\eps_{\lambda'} }    ,
\eeq
\begin{equation}
   G_{n,n^{\prime}}^l(k^2_{\perp},X_{12}) = \int dx_{12}
   \,g_l(k^2_{\perp},x_{12})\,
 y_n(X_{12}{+}x_{12}/2)  y_{n^{\prime}}(X_{12}{-}x_{12}/2),
\end{equation}
where  
$\lambda {=} (n, {\bf k}_{\perp})$,  
$\eps_{\lambda} {=} \eps_n + k^2_{\perp}/2m $,  
$\eps_n,\; y_n $ stand for the energies and wave functions 
of the one-dimensional Schr\"odinger equation, respectively. 
The prime in the sum of eq.~(7) means that the summation
is carried out over ($\lambda, \lambda' $) which are 
not included  in the
model space. It is convenient to put in this sum $\eps_\lambda  {<} 
\eps_{\lambda '} $
(multiplying the result by the factor 2). Then the sum
contains all the states with $\eps_{\lambda}  {>} \mu,\; 
\eps_{\lambda'} {>} 0$.

To overcome a problem of a slow convergence of the integrals of eq.~(7), 
a renormalization can be made in terms of the free $T$-matrix:
\begin{equation}
T={\cal V} + {\cal V} A^0T,
\end{equation}
where $A^0$ is the propagator of two free particles.

Eqs.~(6),~(7) hold true for the $T$-matrix
(with the substitution $T \to \Lambda , B^0 \to B $), but now
the coefficients 
$T_{ij}(X,X';E), B^0_{ij}(X,X';E) $ of the separable
expansion depend only on the difference $t=X-X'$ of the CM coordinates:
\beq
T_{ij}(t;E) =\lambda_{ij} \delta(t) +
\sum_{lm} \lambda_{il}\int dt'\,B_{lm}(t-t';E)
\,\Lambda_{mj}(t';E),
\eeq

The renormalized equation for the effective interaction in a compact
form reads:
\begin{equation}
\Lambda_{ij} = T_{ij} + \sum_{lm} T_{il}\,(B_{lm} - B_{lm}^0)\, T_{mj} 
\end{equation}

   The kernel of this equation converges much faster than 
the original one. Of course, the problem of the slow convergence
at large momenta does not disappear. It passes to eq.~(10), 
but in this case the difficulty can be overcome much more easily.
Indeed, we are dealing now
with the one-dimensional vector $T_{ij}(t) $
instead of the two-dimensional matrix $\Lambda_{ij}(X,X') $.
It is convenient first to find the $T$-matrix in the momentum
representation
by solving the following set of equations:
\begin{equation}
T_{ij}(P_x;E) =\lambda_{ij}  +
\sum_{lm} \lambda_{il}\, B^0_{lm}(P_x;E)
\,T_{mj}(P_x;E),
\end{equation}
where
\begin{equation}
B^0_{lm}(P_x;E) = \int{ d^3{\bf k} \over {(2\pi)^3} }
 {{ g_l(k^2)\,g_m(k^2)} \over {E - {{P_x^2}/{4m}} - {k^2 / m}  }}.
\end{equation}
Then $T_{ij}(t) $ can be found from $T_{ij}(P_x)$ with the help of
the inverse Fourier transformation:
\beq
T_{ij}(t;E) = \int\limits_{-\infty}^{\infty} \frac {dP_x}{2\pi}
\,T_{ij}(P_x;E) \exp (-iP_xt). 
\eeq

The form factors $g_i$ in eq.~(13) are rational functions on $k^2$ 
\cite{Par1,Par2}, namely, combinations of the Yukawa function
and their derivatives with different masses 
$\beta_{in}\; (n=1,\ldots ,4)$.
This integral can be evaluated analytically, but such calculation
is very cumbersome because of a huge number ($\simeq$70)
of particular terms appearing in the integrand. We prefer, following
\cite{BLSZ1}, to integrate it numerically, with the cut-off 
momentum $k_c=60$ fm$^{-1}$ which guarantees an accuracy better than 1\%. 

The Fourier integral (14),
after separating the constant term $\lambda_{ij}$,
 was calculated in \cite{BLSZ1} by direct integration
along the real $P_x$-axis. This method works well at small 
$t$, but for $t > (2 \div 3)\,$fm the integrand contains rapidly 
oscillating factors
multiplied by a slowly falling function $(T_{ij}(P_x^2) - \lambda_{ij}) \simeq 
1/P_x^2 $,  
which makes the convergence very poor. To obtain a reasonable
accuracy at $t{ = } (4 \div 5)$\,fm, the large value of the cut-off momentum 
$P_x^c=3000$\,fm$^{-1}$ was taken in \cite{BLSZ1} with very small 
integration step.
It is much better, in accordance with the general recipe of 
integrating rapidly oscillating functions, 
to integrate eq.~(14) in the complex plane
of $P_x$. The integration contour $C$ can be closed in the upper half-plane
 and deformed to a form convenient for numerical integrating (Fig.~1).
As it can be readily shown, in the case of rational form
factors $g_i$ all the singularities of $T_{ij}(P_x;E)$ are the poles
(simple and multiple) located on the imaginary axis (symmetrically to
the origin). Their position $P_x^{\alpha}  {=} i \gamma_{\alpha} $ 
is defined by various
combinations of the masses $\beta_{in}$ and the $\mu$-dependent
parameter $\gamma_0=\sqrt{-8m\mu}$.  
The contour $C$ surrounds all these poles,
each one yielding a falling exponent $\exp(-\gamma_{\alpha}  |t|)$.
Besides, the procedure of numerical integration of eq.~(13) produces
some additional ``false'' poles depending on $k_c$. Though their contribution
is negligible, the contour
surrounds them also for sake of consistency. In practice, we used
the contour $C$ with values of parameters $a=2$\,fm$^{-1}$, $b=120$\,fm$^{-1}$.
In this case it is placed far enough from all the poles and the integral
can be calculated without numerical problems yielding the correct result
for any value of $t$. 

We describe the procedure of calculating the off-shell $T$-matrix
in detail because, as will be shown, the difference $B-B_0$ in eq.~(11) 
turns out to be to be rather small and the solution
of this equation with a good accuracy coincides with the $T$-matrix.
Therefore the evaluation of this quantity is of primary importance.

In this paper, we go from the semi-infinite system to 
a more realistic geometry of a finite slab within the Saxon-Woods
potential well. The parameters are chosen in such a way to reproduce
qualitatively those of heavy nuclei in the lead region:
the width of the slab is $2L=16$\,fm, the 
depth of the well $V_0=-50$\,MeV and the diffuseness parameter $d=0.65$\,fm.
In the slab case, all the above equations are valid
with substitution of the slab wave functions and energies in eqs.~(7),\,(8).
Instead of direct solution of these equations, the
Local Potential Approximation (LPA) is used which was previously proved
to be a reliable approximation
for semi-infinite nuclear matter \cite{BLSZ1}. 
Within the LPA, the exact
values of $B_{lm}(X_1,X_2;E)$ are replaced by the set
of those $B_{lm}^{\mbox{\scriptsize inf}}(t,E;V[X])$ for 
infinite nuclear matter put in the potential well $V[X]$. 
Here $X=(X_1+X_2)/2$ is the average value of the two CM coordinates.
As it is shown in \cite{BLSZ1}, comparing exact
solution for semi-infinite matter 
with the LPA prediction, the latter works with an accuracy of  
few percent even in the surface region. It is natural
to suppose that in the slab case the LPA works also sufficiently well.  
The LPA procedure is as follows. 
At a fixed value of the chemical potential $\mu$,
it is necessary, first, to calculate the set of vectors  
$B_{lm}^{\mbox{\scriptsize inf}}(t,E=2\mu;V_i)\; 
(V_i{=}\delta V {\cdot} (i-1))$, and then,
for every value of $(X,t)$, to
find $B_{lm}^{\mbox{\scriptsize LPA}}(X_1,X_2)$ by interpolating 
values of $B_{lm}^{\mbox{\scriptsize inf}}(t;V_i)$ with nearest to
$V(X)$ values of $V_i$.

Let us first take $\mu{=}-8$\,MeV.
Shown in Fig.~2 is the difference 
$(B_{lm}^{\mbox{\scriptsize inf}}(V) - B_{lm}^0)$ for three main components 
$lm$, in comparison with the free propagators  $B_{lm}^0$,
for the case of the maximum value of the potential depth $V=50$\,MeV. 
Of course, for smaller values of $V$ this difference is even less.
To analyse the
origin of this smallness, let us separate the sum of eq.~(7)
into two parts, the first one with both positive energies and the second
one with one negative energy, and compare each of them with the
corresponding sum for the free propagator. 
It should be stressed that within the LPA eq.~(7) is very similar
to that for $B_{lm}^0$ because  the plane waves
in the constant potential $V_0{=}V(X)$ stand for $y_n$ in this case. 
Every term of the first sum corresponds
directly to the one of $B_{lm}^0$ with 
the same energy denominator. It can be easily seen that
the numerator of the in-matter term is less than that
of the free one because the matrix elements of the form factors
are calculated for bigger values of momenta ($q=\sqrt{p^2+2mV}$ 
in the in-matter case in comparison with $p$ in the free one).
Since all the form factors $g_i(p^2)$ fall with $p$, this part of  
$B_{lm}^{\mbox{\scriptsize inf}}(V)$ (in absolute value) is less 
than $B_{lm}^0$
(at $20 \div 30$ \%). The second part of the sum under consideration
has no analogue in the free propagator. Though it  
contains an essentially smaller phase space that the first one,
its value turns out to be significant due to small value of the energy 
denominators.
Its contribution to the absolute value of 
$B_{lm}^{\mbox{\scriptsize inf}}(V)$ 
is even larger than the difference of the first term of
$B_{lm}$ and $ B^0_{lm} $, so the absolute
value of $B_{lm}$ exceeds that of $B^0_{lm} $, but only a bit.
The size of the difference is about 10\%. 

Fig.~3 shows similar comparison of the effective interaction 
${\cal V}^{\mbox{\scriptsize inf}}_{\mbox{\scriptsize eff}}\, 
[V{=}50$\,MeV] and the free T-matrix. Of course, the
$\delta$-terms of both amplitudes (see eqs.~(6), (10))
are extracted. For more descriptive comparison
of the ``exact'' LPA effective interaction and the free $T$-matrix 
we display in Fig.~4 their zero-order moments:
\begin{equation}
\bar{\Lambda}_{ij}(X) = \int dt \,\Lambda_{ij} (X{-}t/2,X{+}t/2),
\end{equation}
and the same for $\bar{T}_{ij}$.
As it is seen, the difference between these averages 
is again very small for the 11-component, but for other components
it is greater, namely $\simeq 20$\% for the 12-component and 
$\simeq 30$\% for the 22-one. However, these two deviations are
of the opposite sign and, as we show just below, almost
compensate each other in the sum  at values of $k^2, k'^2$
which are important for the pairing problem.

It can be demonstrated by analysing the localized form of the
effective interaction which was introduced in \cite{BLSZ1}: 
\begin{equation}
\bar {\cal V}^F_{\mbox{\scriptsize eff}} (X)=
\sum_{ij} \bar {\Lambda}_{ij}(X)
g_i(k^2_F(X)) g_j(k^2_F(X)),
\end{equation}
where $k^2_F(X){=} 2m(\mu{-}U(X))$ if $\mu{-}U(X)>0$ and $k^2_F(X){=}0$
in the opposite case.
In \cite{BLSZ1}, it is argued that, for states in the Fermi surface
vicinity, the effective interaction can be approximately replaced by
$\bar {\cal V}^F_{\mbox{\scriptsize eff}} (X)$.
It is drawn in Fig.~5, together with $\bar T^F$ which is defined
in a similar way. One sees that the compensation discussed above, 
indeed, occurs and the difference between the two curves is 
practically negligible.

To imitate the situation in the vicinity of drip-lines,
all the calculations were repeated for $\mu{=}-4$MeV. 
Results are completely similar to those for $\mu{=} -8$MeV.
They are shown in Fig.~6 for
the $B_{11}$ component and in Fig.~5, for the local form of the 
effective interaction 
$\bar {\cal V}^F_{\mbox{\scriptsize eff}}(X)$. 
Thus, this simple approximation of $\Veff$ with the free
$T$-matrix can be used also for predicting properties of nuclei
nearby the drip-lines. It should be mentioned that one approximation
used in this analysis looks doubtful for small values
of $\mu$. This is neglecting all the pairing effects in the
complementary space. This drawback can be improved
by small change of the model space by
addition to it a set of states with small positive
energies $\eps_{\lambda} < E_0$ where $E_0$ is of the order
of several MeV. Estimates show that in this case the difference
between $\Veff$ and the $T$-matrix becomes even smaller.   

Thus, the free off-shell $T$-matrix taken 
at the energy $E=2\mu$, indeed, is a good approximation for 
the microscopic effective pairing interaction.
For the separable representation of the Paris $NN$-potential 
it can be easily found by solving eqs.~(11)--(13). As far as we
deal with comparatively small shift from the mass shell,
all the realistic $NN$-potentials must give approximately
the same predictions for the $T$-matrix.

\vskip 0.3 cm
Two of the authors (E.S. and M.Z.) thank 
INFN and Catania University for hospitality during their stay
in Catania. One of us (E.S.) thanks also the Italian Ministry of
Foreign Affairs and the Landau Network-Centro Volta for support
during the autumn of 1999 when this work was carried out. M.Z.
thanks Peter Schuck and Nguyen Van Giai for the warm hospitality 
at IPN Orsay.

\newpage 
\begin{figure} 
\epsfxsize=16.5cm
\epsfysize=22.cm
\vspace*{-4 cm}
\centerline{\epsffile{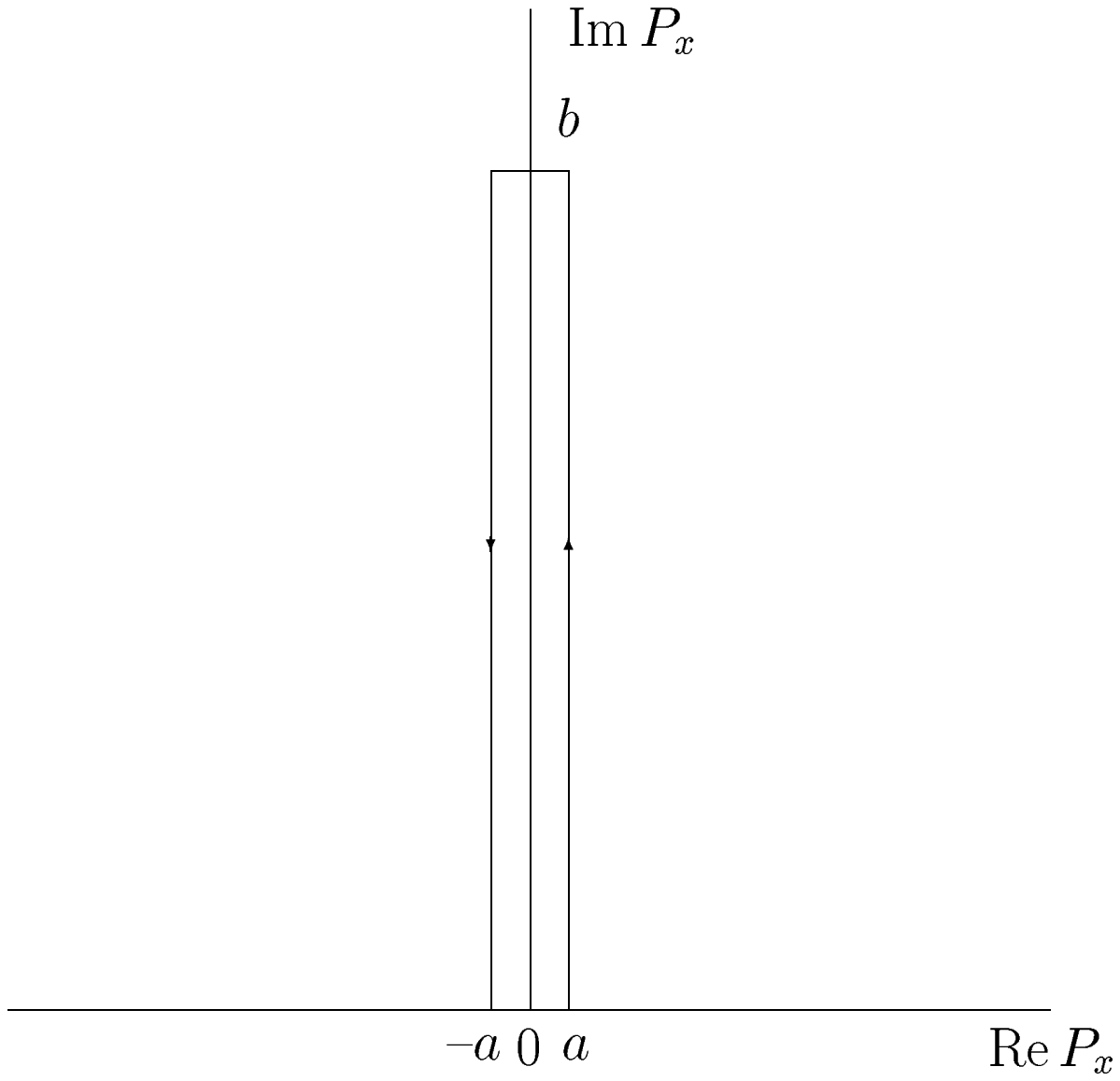}}
\vskip -7 cm
\caption[]{
The integration contour for the inverse Fourier transformation (14).
}
\end{figure}

\newpage 
\begin{figure} 
\epsfxsize=16.5cm
\epsfysize=22.cm
\centerline{\epsffile{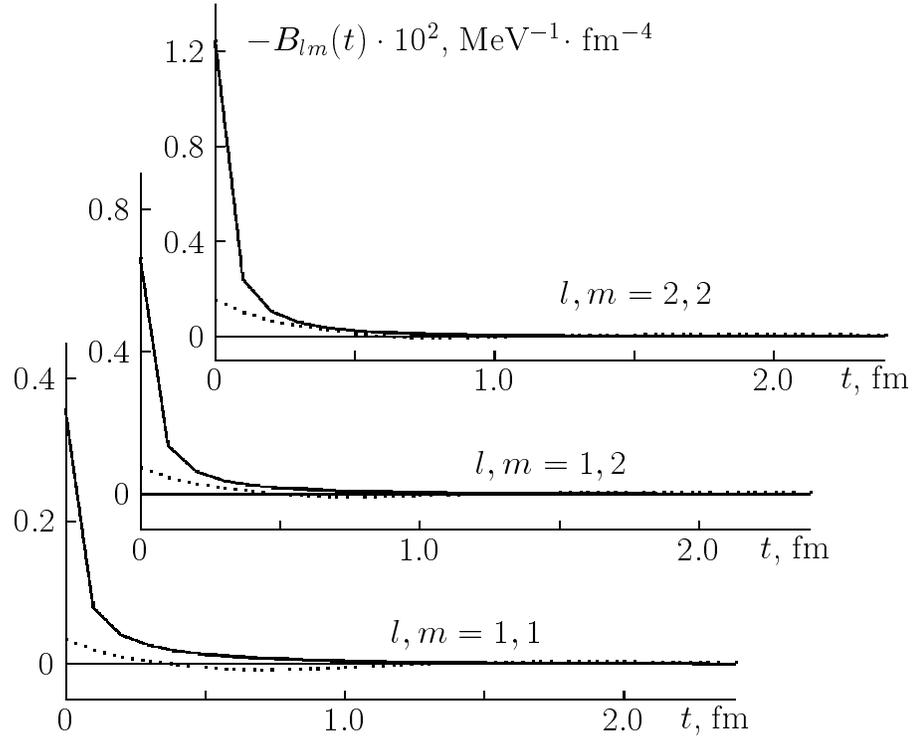}}
\vskip -7 cm
\caption[]{
The free propagators $-B^0_{lm}$ (solid lines) and the differences
$-(B_{lm}^{\mbox{\scriptsize inf}}[V=50\,\mbox{MeV})]-B^0_{lm})$
(dotted lines) calculated for $\mu=-8$\,MeV versus relative
coordinate $t=X-X'$. 
}
\end{figure}

\newpage 
\begin{figure} 
\epsfxsize=16.5cm
\epsfysize=22.cm
\centerline{\epsffile{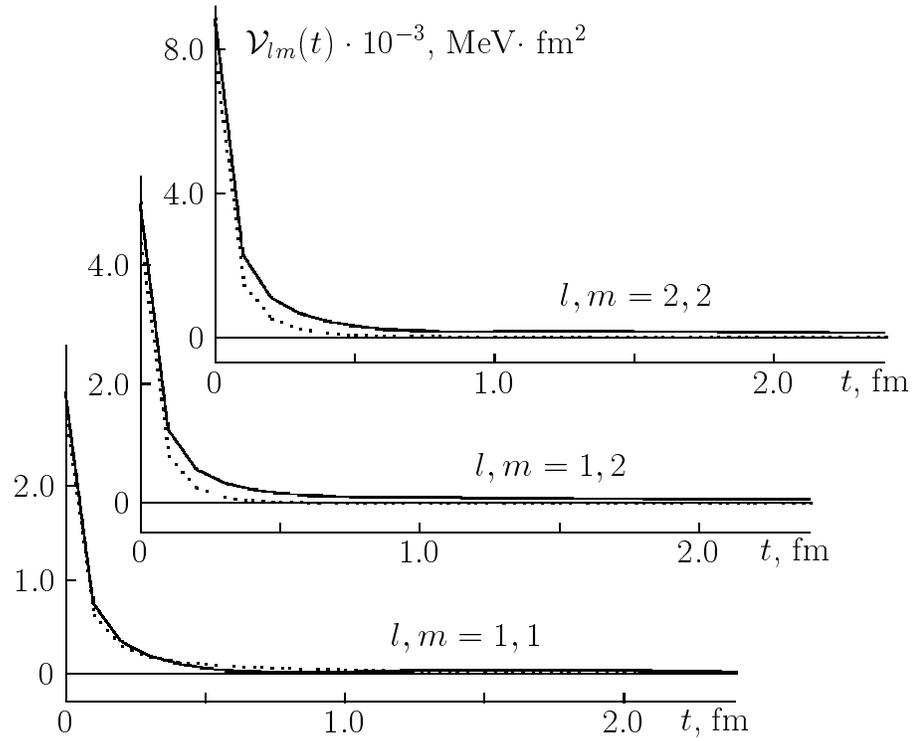}}
\vskip -7 cm
\caption[]{
The $lm$-components of the effective interaction
${\cal V}_{\mbox{\scriptsize eff}}^{\mbox{\scriptsize inf}}[V=50\,\mbox{MeV}]$
(solid lines) and the free $T$-matrix (dotted lines) calculated
for $\mu=-8\,$MeV versus $t$. Sign is changed for $i,j=1,1$ and $2,2$.
}
\end{figure}

\newpage 
\begin{figure} 
\epsfxsize=16.5cm
\epsfysize=22.cm
\centerline{\epsffile{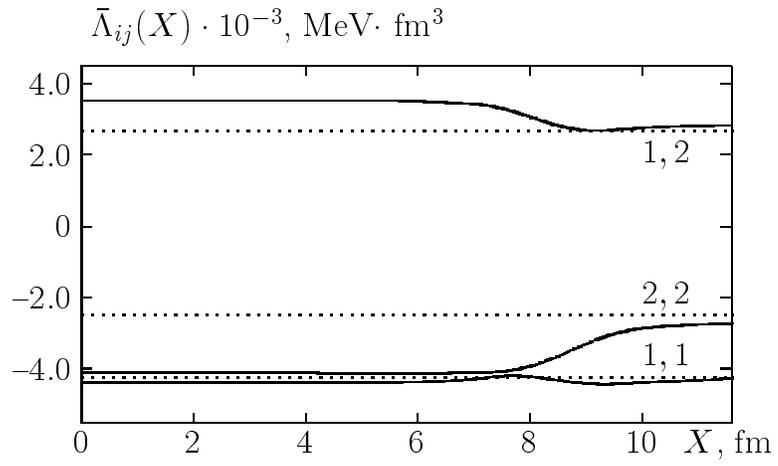}}
\vskip -7 cm
\caption[]{
The zero moments $\bar\Lambda_{ij}(X)$ (solid lines) and the values
$\bar T_{ij}$ (dotted lines) for $\mu=-8$\,MeV.
}
\end{figure}

\newpage 
\begin{figure} 
\epsfxsize=16.5cm
\epsfysize=22.cm
\centerline{\epsffile{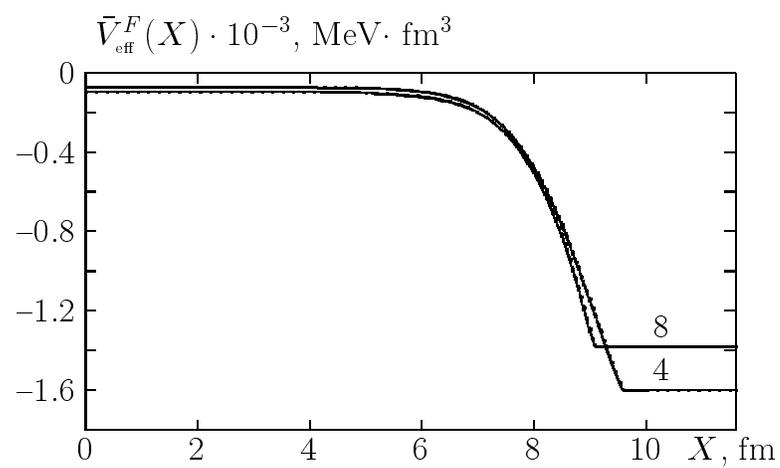}}
\vskip -7 cm
\caption[]{
The effective interaction $\bar {\cal V}^F_{\mbox{\scriptsize eff}}(X)$
(solid lines) and $\bar T^F_{ij}(X)$ (dotted lines) for two values
of the chemical potential $\mu$ (the number corresponds to $|\mu|$ 
in MeV). 
}
\end{figure}

\newpage 
\begin{figure} 
\epsfxsize=16.5cm
\epsfysize=22.cm
\centerline{\epsffile{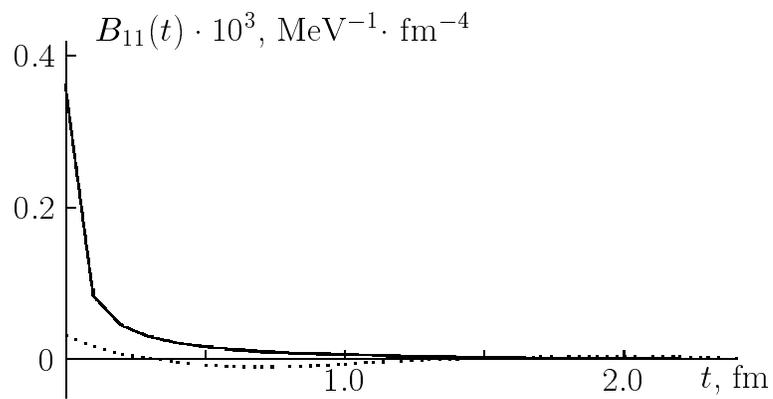}}
\vskip -7 cm
\caption[]{
The same as in Fig.~2 for $l,m=1,1$ calculated for $\mu=-4$\,MeV.
}
\end{figure}

\end{document}